# Transfer Learning Based Multi-Objective Genetic Algorithm for Dynamic Community Detection

Jungang Zou[a,b], Fan Lin[a,*], Siyu Gao[a,d], Gaoshan Deng[a],

Wenhua Zeng[a], Gil Alterovitz[c]

[a]Department of Software Engineering, Xiamen University, Xiamen, 361005, Fujian, China
[b]Department of Biostatistics, Columbia University, New York, 10032, NY, USA
[c]Biomedical Cybernetics Laboratory, Harvard Medical School, Boston, 02115, MA, USA
[d]Carnegie Mellon University, Pittsburgh, 15213, PA, USA

## Abstract

Dynamic community detection is the hotspot and basic problem of complex network and artificial intelligence research in recent years. It is necessary to maximize the accuracy of clustering as the network structure changes, but also to minimize the two consecutive clustering differences between the two results. There is a trade-off relationship between these two objectives. In this paper, we propose a Feature Transfer Based Multi-Objective Optimization Genetic Algorithm (TMOGA) based on transfer learning and traditional multi-objective evolutionary algorithm framework. The main idea is to extract stable features from past community structures, retain valuable feature information, and integrate this feature information into current optimization processes to improve the evolutionary algorithms. Additionally, a new theoretical framework is proposed in this paper to analyze community detection problem based on information theory. Then, we exploit this framework to prove the rationality of TMOGA. Finally, the experimental results show that our algorithm can achieve better clustering effects compared with the state-of-the-art dynamic network community detection algorithms in diverse test problems.

\* Corresponding author. *E-mail address:* iamafan@xmu.edu.cn .

## 1. Introduction

Community detection problem, a fundamental research area in artificial intelligence and network science, can be widely applied to solve problems in ad-hoc networks, social networks [1], communication networks, and scientific cooperation networks, etc. It needs to divide networks into several substructures named community (clustering) whose internal nodes are closely connected. The association between the nodes within the same community is tight, while the correlations between different communities are relatively sparse. The community is essentially a strong association subset of the entire network. Although community detection is rather important, however, the static network community detection problem has been proved to be a N-P problem [2].

In real life, many network structures are actually dynamic, such as dynamic social network, dynamic protein-protein interaction networks, dynamic energy consumption networks. The community detection problem on dynamic network can also be viewed as a set of static network community detection problems, thus more difficult and complicated. Fortunately, a framework called *temporal smoothness* has been introduced to solve the problem of dynamic community detection, which assumes the dramatic changes of communities are not expected in a short period of time, so each community is smoothed over time [3, 4, 5, 6]. In such framework, it makes a trade-off between two different criteria to realize smoothness. One is to maximize clustering accuracy on each snapshot and the second objective which is to minimize the difference of clustering results between two continuous time snapshots should also be considered. In another word, we have to achieve a desirable partition quality on the current time step and simultaneously make the clustering results between two time points as consistent as possible, keeping the stability of the communities in the whole system. These two objectives are conflict some time, thus form the problem of multi-objective optimization. To solve that, the multi-objective evolutionary algorithm has been widely used and has achieved fairly good performance [3, 7, 8].

We believe that dynamic network models do not change randomly. On the contrary, some communities still remain some of the invariant features of the previous snapshot in the process of evolution. If we can identify and extract these features, we are able to provide helpful information to the following optimization process. Inspired by the migration of feature transfer learning [9], we introduce this idea of feature transfer to solve the dynamic network problem at the first time. Then, we build a set of mechanisms for extracting features from the dynamic community network structure and applying these features to obtain better solutions. We combine these feature migration mechanisms with the classic multi-objective evolutionary algorithm NSGA-II [10], and propose a multi-objective evolutionary algorithm based on feature transfer (feature Transfer based Multi-objective Optimization Genetic Algorithm, TMOGA), which significantly improves the performance of the original NSGA-II algorithm in solving the dynamic community problem, in both accuracy and efficiency.

Additionally, although the theoretical analysis on community detection problem has been widely researched in network science, it strongly depends on the data and specific evaluation criteria. To deal with these inconveniences, we propose a new theoretical analysis framework based on information theory. From the informational perspective, community detection can be regarded as information compression process, which refines information from node level to subgraph level. Different clustering criteria corresponds to different mechanism compressing information. With this idea, we can unify different evaluation criteria for community detection into a uniform scheme to avoid the inconvenience when analyzing. Moreover, in this informational framework, the rationality of feature transfer discussed above can be easily proved from the view of information compression.

The main contributions of this paper are as follows:

1. Based on NSGA-II framework, we propose a novel dynamic network community detection algorithm based on transfer learning mechanism. We then implement the details of feature transfer mechanism, including the feature extraction step to discover cliques, and the feature migration step transferring features to next initial populations.

2. For the first time, we propose a new framework for theoretical analysis on community detection problem based on information theory. By introducing *Information Bottleneck* model [11], we successfully abstract community discovery into the process of information compression. After that, different criteria for clustering can be unified into an analyzing framework. Based on it, the rationality of feature transfer can be proved straightforwardly.

3. We verified the effect of the TMOGA algorithm by a sequence of experiments. The experimental results show TMOGA has advantages in clustering accuracy and convergence time in classic test sets, practical problems, and even initialization phrase, when compare to other similar algorithms on the forefront.

## 2. Related Work

Compared with the mature research of static community detection, the research of dynamic community detection is a relatively new research direction. Traditionally, methods proposed to solve this problem depend on mathematical characteristics of dynamic network such as [12, 13, 14, 15]. FacetNet [13], one of the most classic algorithms in them, used matrix-based method to capture communities with a fast speed.

In 2013, the multi-objective optimization algorithm has been gradually introduced into the problem of community detection. Based on NSGA-II, the most famous multi-objective optimization algorithm, DYNMOGA [3] proposed by Folino, achieved the state-of-art effect in that time. DYNMOGA regards dynamic community detection as a multi-objective problem under the framework of *temporal smoothness* and utilizes genetic algorithm to obtain the best partitions. After its success, multiple modified algorithms have been proposed: DYNMODPSO [8] exploits particle swarm evolutionary algorithm to ensure accuracy of optimization by random walk initialization, while a correction mechanism proposed can improve the solutions at the first two snapshots; DECS [7], on the other hand, identifies all inter-edges in communities and out-edges connected with communities. It directly encodes solutions on adjacency matrix and uses an edge migration operation to enhance results.

However, these advanced algorithms only improve on the optimization at current network while information from solutions at last network has been ignored. Inspired by the success of transfer learning in evolutionary algorithms [16, 17, 18, 19, 20], we consider it for dynamic community detection. As a new machine learning method, the core idea of transfer learning is to solve problems in different but related fields through using existing knowledge. In clustering problem, transfer learning was used by Dai [21] for clustering analysis for the first time. However, there is no research on transfer learning for dynamic community detection at present.

Therefore, we applied transfer learning and a multi-objective evolutionary algorithm to dynamic community detection, and the quality of the evolutionary algorithm is improved while meeting multiple objectives.

## 3. Problem Description

The goal of dynamic community detection is to find the community structure at each time in a dynamically changing community network structure. Define $N = \{N^1, N^2, \ldots, N^T\}$ as the sequence of dynamic networks and each $N^t$ corresponds to a network at snapshot $t$. Let $V^t$ and $E^t$ denote the collections of nodes and edges in network $N^t$. Then, the community in a static network $N^t$ can be expressed as a subgraph where it has high edge density inside. We define $CR^t = \{C_1^t, C_2^t, \ldots, C_k^t\}$ as the discovered communities, and the element $C_i^t$ is a subgraph of $i$-th community. Since our algorithms can only identify the non-overlapping community structure, the constrains should hold: $C_1^t \cup C_2^t \ldots \cup C_k^t = N^t$ and $C_1^t \cap C_2^t \ldots \cap C_k^t = \emptyset$. To make it clear, we define "neighbors" of a node which have edge connecting with it.

### 3.1. Temporal Smoothness

From what has been discussed in section 1, we expect the structures of dynamic networks are not dramatically changed over time. To model this evolving pattern, a framework called *temporal smoothness* has been proposed [3, 4, 5, 6]. In this framework, the algorithms should find the best community structures at each network $N^t$ and also balance the similarity degree between communities of adjacent time. Thus, a quantity named snapshot cost (*SC*) and a quantity called temporal cost (*TC*) are used to model the evaluations of community structures and the similarity respectively. As a result, the total cost of dynamic community detection will be $\alpha * SC + (1 - \alpha) * TC$, $\alpha$ is some predefined coefficient between [0, 1], to balance *SC* and *TC*. However, α is difficult to specify in distinct context, so we treat that as multi-objective optimization problem and these 2 objectives are both minimized simultaneously in algorithms.

### 3.2. Evaluation Criteria for SC

Community detection problems can also be defined as a clustering problem, several criteria have been proposed to evaluate it, e.g. *Modularity* [22], *Community Score* [23], etc. In our algorithm, *Modularity*, also known as *Q*-function, will be applied to evaluate the snapshot cost. It can be expressed as the following mathematical form:

$$Q(CR^t) = \sum_{i=1}^{|CR^t|} \left[ \frac{l_i}{|E|} - \left( \frac{d_i}{2|E|} \right)^2 \right] \tag{1}$$

Where $|E|$ is the number of edges at $N^t$, $l_i$ represents number of edges inside $i$-th community and $d_i$ denotes the total number of degree in $i$-th community. The value range of *Modularity* is [−1/2,1]. The closer to 1 the value is, the stronger the strength of the community structure. In this case, the solution with higher *Modularity* has greater community structures and lower snapshot cost, i.e. $SC(CR^t) = -Q(CR^t)$.

### 3.3. Evaluation Criteria for TC

In addition to the intensity of community nature, there is also a need to measure *TC*. We use the *Normalized Mutual Information (NMI)* [24]. It is defined on discrete random variables as follow:

$$NMI(CR^{t-1}; CR^t) = \frac{-2 \sum_{i=1}^{|CR^t|} \sum_{j=1}^{|CR^{t-1}|} C_{ij} \, log\left( \frac{C_{ij}|V|}{|C_i||C_j|} \right)}{\sum_{i=1}^{|CR^t|} |C_i| \, log(|C_i|/|V|) + \sum_{j=1}^{|CR^{t-1}|} |C_j| \, log(|C_j|/|V|)} \tag{2}$$

Where $C$ is the confusion matrix [25], $C_{ij}$ presents the number of common nodes of community $i$ in partition $CR^{t-1}$ and community $j$ in partition $CR^t$, $|CR^t|$ expresses the total number of communities in partition $CR^t$, $|C_i|$ is the number of nodes in community $i$, and $|V|$ is the number of nodes at $N^t$. *NMI* ranges from [0, 1]. $NMI(A; B) = 1$ when $A = B$; $NMI(A; B) = 0$ when $A$ is completely different from $B$. The greater *NMI* is, the two community structures of adjacent time have lower temporal cost, i.e. $TC(CR^{t-1}; CR^t) = -NMI(CR^{t-1}; CR^t)$.

*3.4. Problem Definition*

From what we have discussed above, because we need to minimize *SC* and *TC* at the same time on dynamic network, the problem can be defined as a multi-objective optimization problem:

$$\underset{CR^t}{argmin}\begin{cases} -Q(CR^t), t \geq 1 \\ -NMI(CR^t; CR^{t-1}), t \geq 2 \end{cases} \qquad (3)$$

For instance, we assume that we have a simple model for dynamic network community detection problems in Fig.1, as described below $T = 2$, $N = \{N^1, N^2\}$. In this case, we need to find a best partition $CR^2$ minimizing function (3) simultaneously.

*3.5. Multi-objective Optimization Algorithm*

The most famous algorithm in the field of multi-objective optimization is the NSGA-II [10] algorithm proposed by Deb in 2002. It is a genetic algorithm and sorts solutions by its Pareto rank [26], where each solution has comparative advantage to others in the same rank in one objective direction, and is better than at least one solution in all objective directions in next rank. An example is reported in Fig.2, where each dot corresponds to a solution. Finally, it returns a set of Pareto solution (solutions at first rank). Also, NSGA-II adopts elitist strategy which takes crossover operation only among the individuals with high fitness. Compared with previous algorithms, NSGA-II consists of several methods to calculate Pareto rank with much lower complexity. Although further development of NSGA-II has been proposed like NSGA-III [27], they have higher efficiency in problems with more than 5 objectives but slower in low dimensional optimization compared with NSGA-II. Thus, NSGA-II is specified to optimize objective functions (1), (2) in our algorithm.

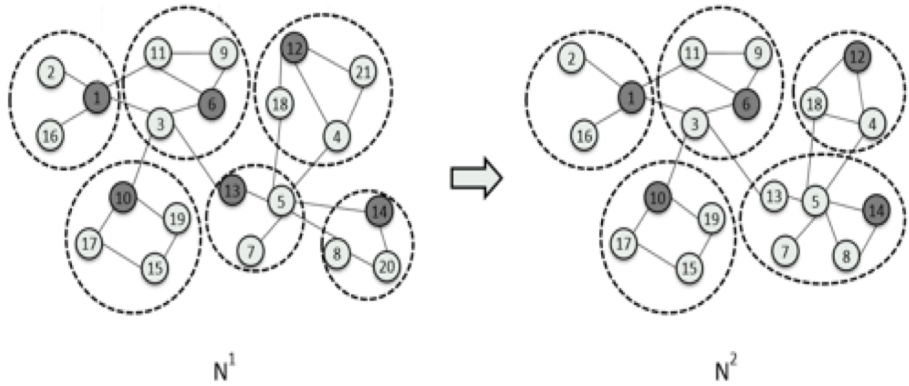

Fig. 1. A simple dynamic network model of dynamic networks

*3.6. Evaluation Criteria for Pareto Solutions Selection*

After the process of genetic algorithm NSGA-II, a set of Pareto solutions will be eventually obtained. To select the best solution from them, the criteria *Community Score* of order 2 is applied to sort these solutions. *Community Score* measures the quality of the community structure results. The mathematical definition of it is as follow:

$$CS(CR^t) = \sum_{i=1}^{k} \left[ \frac{\sum_{m \in C_i^t} (\mu_m)^2}{|C_i^t|} \times \sum_{m,n \in C_i^t} A_{mn} \right] \tag{4}$$

Where $\mu_m = \frac{1}{|C_i^t|} \sum_{n \in C_i^t} A_{mn}$, $|C_i^t|$ denotes the number of nodes in community *i* at time point *t* and $A$ is the adjacency matrix. The larger the value of *Community Score* is, the better community partition we obtain. In our algorithm, we thus select the solution with highest *Community Score* as the best among all Pareto solutions.

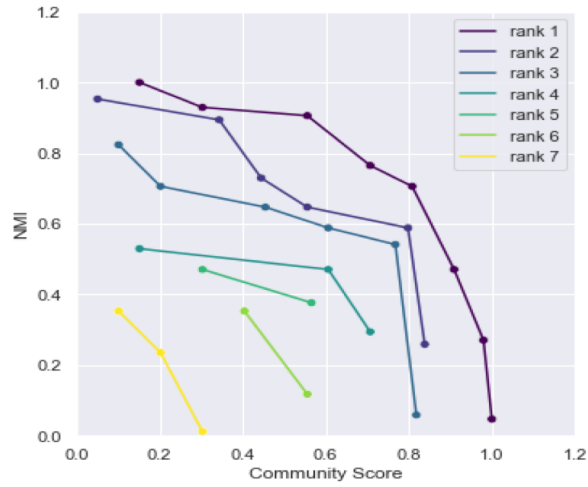

Fig. 2. Example of Pareto rank, each dot is a solution

# 4. Feature Transfer

In this chapter, a mechanism called feature transfer is introduced to improve the results of dynamic community detection.

*4.1. Transfer Learning in Dynamic Community Detection*

Transfer learning, a popular areas of machine learning, overcomes the requirements of traditional machine learning where training sets and test sets are identically distributed. The goal of transfer learning is to make a model utilizing the knowledge learned in the past to assist with solving new problems. In the case of a new problem with a strong correlation with an old problem, transfer learning has great efficiency.

For traditional algorithms for solving dynamic problems [3], each moment is restarted with random initialization in order to optimize the current network. This kind of optimization often discards the previous results directly, and independently optimizes the network at each snapshot. In this sense, we can apply transfer learning mechanism to generate better initial solutions by utilizing the result obtained in last snapshot, to accelerate the convergence of genetic algorithm. Generally, there are mostly 2 types of transfer learning: sample transfer and feature transfer. We will discuss their applicability then.

## 4.2. Sample Transfer

Sample transfer is based on the instance object under the isomorphic space which sets the initial solution of current snapshot the same as the results from previous network completely. However, in the context of dynamic network communities, since the structures at each moment change, therefore it is likely to meet overfitting if we directly set initial solutions the same as the results from last snapshot. Due to this limitation, sample transfer is not suitable for coping with dynamic network community detection problem.

## 4.3. Feature Transfer

On the other hand, feature transfer exploits the features of objects to enhance the result of clustering. The rationality of feature transfer arises from the fact that there still are some intersections between the two very different graphs at the subgraph level. Therefore, the feature transfer is a much widely useful method for dynamic networks. An intuitive illustration for the pattern of feature transfer is in figure 3.

**Feature Definition** In dynamic networks, the inheritable features should have the characteristics associated with the network community detection characteristics. Therefore, we consider the use of "small clique" as a feature of the graph structure. Small clique is essentially a subset of nodes belonging to a community, and a community structure may contain multiple small cliques. The definition of a small clique is a set of nodes within a community whose clustering effect exceeds a certain threshold and threshold can be *Community Internal Density* (*CID*):

$$CID(S) = \frac{2L(S)}{|S| * (|S| - 1)} \tag{5}$$

Where $S$ represents any subgraph, $L(S) = \sum_{j,k \in S} A_{jk}$ , $A$ is the adjacency matrix, $|S|$ indicates the number of nodes in the subgraph. When the threshold is set as 0, the whole subgraph is a feature; when the threshold reaches 1, only fully-connected subgraphs represent features.

**Advantages of Feature Transfer** Compared with sample transfer, the advantage of feature-based transfer is that it balances efficiency and robustness. In the dynamic model, as long as the structure of the graph changes, the overall migration of the samples will have a certain negative impact due to the dissimilar distributions among networks. From a pyramidal perspective, a network is split into multiple communities, a community is split into features and the features consist of nodes. In this pyramid, feature located at a middle level, which has abilities for transfer learning both efficiently and robustly.

## 5. Transfer Learning Based Multi-objective Evolutionary Algorithm

This section introduces the TMOGA algorithm. TMOGA algorithm selectively migrates the characteristics of the network at the previous snapshot to the initial population at the next time point, whose faster convergence rate is adapted to the large-scale data set and the online dynamic community detection problem. Since there is no previous solution which can be used for the feature transfer at $t = 1$, TMOGA uses the single optimization algorithm to generate the final solution of $t = 1$, and then the feature extraction and selection are carried out for the solution of $t > 1$. The mechanism of feature transfer will be separated into two steps, feature extraction and feature migration. In feature extraction step, the algorithm will discover all suitable cliques in communities; in feature migration step, the cliques extracted will be migrated into the initial population in next network. The whole procedure is listed in figure 4.

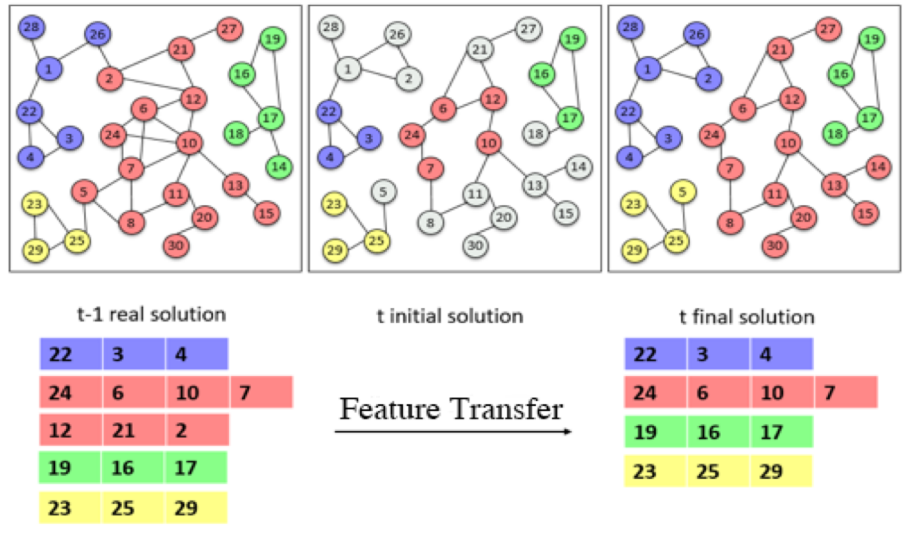

Fig. 3. Feature transfer based on cliques

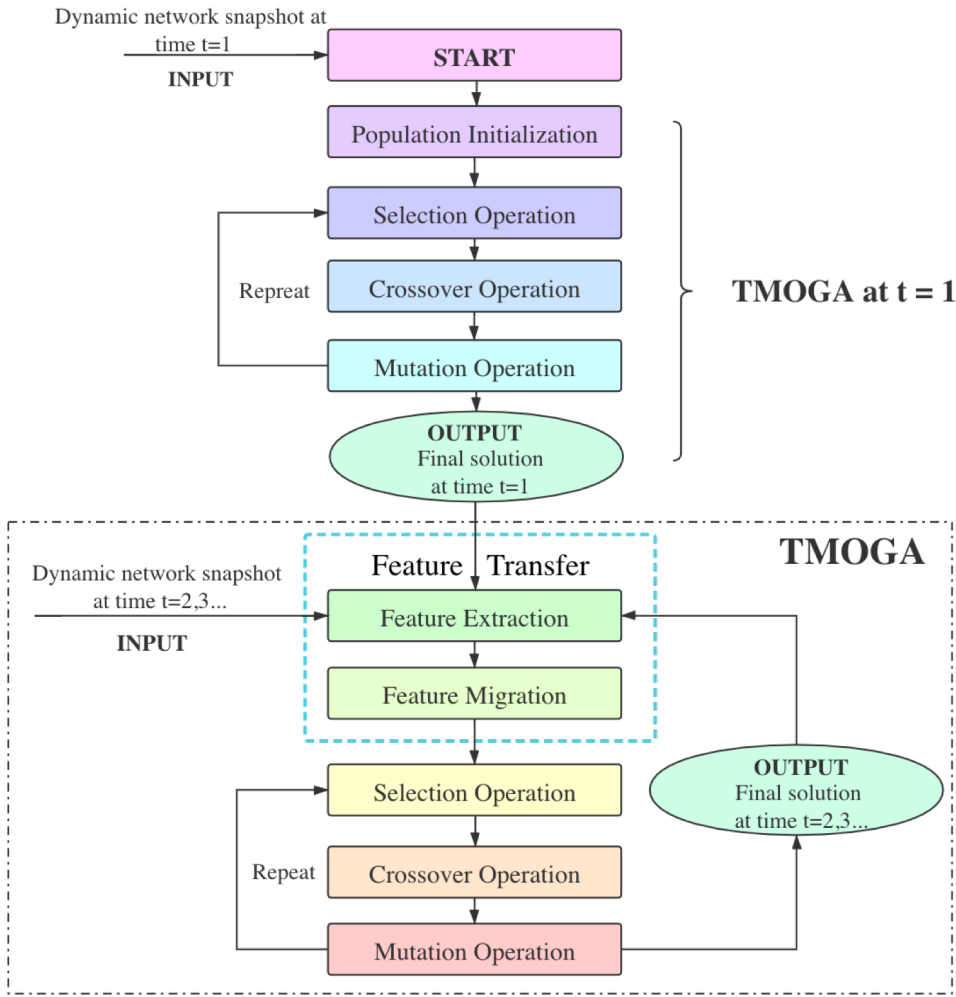

Fig. 4.  TMOGA algorithm framework

## 5.1. Implementation of Feature Extraction (Clique Discovery)

Through the feature extraction method, a community can be divided into a number of small groups called features (or cliques). We represent the coarseness of dividing a community into several cliques by definition of *Community Internal Density (CID)* in section 4.3. When *CID* threshold is set to 0, every community is a small clique; and closer to 1 *CID* threshold is, only the more fully connected subgraphs in community constitute features. We pick up the best solution at the previous time point, and pick out the small cliques existing in it. In figure 5, assuming the graph is classified into a community, then {1, 2, 3, 4} and {5, 6, 7} constitute two small cliques respectively, characterized by tight internal link.

To discover small cliques inside communities, an immediate idea is to enumerate all possible subgraphs exceeds *CID* threshold, then match each desired subgraph to communities. Formally, this idea can be regarded as subgraph isomorphism problem [28]. However, current algorithms to solve subgraph isomorphism problem such as VF2 [29] are too time-consuming. Thus, we propose a new algorithm based on backtracking algorithm with 3 techniques to speed it up dramatically, i.e. greedy pruning, orderly search and depth limit.

As we all know, backtracking algorithm builds a depth-first search tree to find solutions recursively. However, without any improvement, the complexity will become $O\big(|C_i|^{|C_i|}\big)$, $|C_i|$ is the number of nodes inside *i*-th community. To reduce the complexity, we use greedy pruning that stops searching from current subtree if its *CID* is less than the threshold. This pruning technique omits the possible solutions in deeper tree levels and is thus greedy algorithm. Also, since the nodes in any subgraph has no order, we can utilize orderly search that only expand the searching tree to the neighbors whose index is greater than current search node. This technique will reduce the number of subtrees in each tree level and their depth to half on average, which significantly decreases the complexity to $O\left(\frac{|C_i|}{2}^{\frac{|C_i|}{2}}\right)$, Finally, large cliques have larger possibility to change its structure in next network in practice and will lead to overfitting if we simply transfer these large cliques. Therefore, we set the max depth (*Md*) for the search tree to limit the size of cliques. This technique will also limit the huge search space and reduce the search complexity to $O\left(\frac{|C_i|}{2}^{\min\left(\frac{|C_i|}{2},Md\right)}\right)$ in the worst situation, i.e. the fully connected community. Totally, the complexity for clique discovery in all communities will be $O\left(\sum_i^k\left[\frac{|C_i|}{2}^{\min\left(\frac{|C_i|}{2},Md\right)}\right]\right)$, $k$ is the total number of communities. In practice, to keep the characteristic of small cliques and prevent from overfitting, *Md* should be set small. For instance, in social network, due to the famous "six degree of separation" where each node in network can visit other node in less than about 6 steps, *Md* can be set smaller than 6.

For real dynamic networks, the famous small-world effect [30] will continue to improve the efficiency of algorithm evidently. Due to the scale-free effect [31], the degree of nodes follows a power-law distribution and very few nodes have large number of neighbors; this can reduce the width of search tree at each level. Let $d$ denote the average degree in network, the complexity becomes $O\left(\sum_i^k\left[min\left(\frac{|C_i|}{2},\frac{d}{2}\right)^{\min\left(\frac{|C_i|}{2},Md\right)}\right]\right)$. Even in large scale network (*d* is

always less than 1000), $d$ and $Md$ are far less than $|C_i|$, so the total complexity will eventually be $O\left(k\left(\frac{d}{2}\right)^{Md}\right)$. This complexity is notably smaller than the original complexity $O\left(\sum_i^k |C_i|^{|C_i|}\right)$.

After illustrating the techniques to reduce complexity, the whole procedure of feature extraction algorithm can be stated as follows: we first initialize the searching process from a starting node, and then attempt to expand the search tree to neighbor nodes with larger index in community. The expansion of search tree will continue recursively unless there is no neighbor node with larger index or the total depth of search tree exceeds $Md$ or the clique formed by searched nodes has smaller *CID* than threshold. After the searching process from the starting node is completed, a neighbor node with larger index will be selected as starting node and the searching process will repeat respectively. Finally, all small cliques in this community will be obtained. More details can be found in Algorithm 1.

---

**Algorithm 1** Feature Extraction (Clique Discovery)

---

**Input:** Network $N$, community $C_i = \{V_j | V_j \in community\ i\}$, threshold $CID\_threshold \in [0,1]$, max depth $Md \in \{1,2,3 \dots\}$.

**Output:** Cliques set $CL = \{clique_1, clique_2 \dots\}$

1:    **if** size($C_i$) $\leq 2$ **then**
2:       **return** $\{\}$
3:    **end if**
4:    **if** CID($N, C_i$) $\geq CID\_threshold$ **then**
5:       **return** $\{C_i\}$
6:    **end if**
7:    Sort $C_i$ by ascending order according to node index
8:    $CL \leftarrow \{\}$
9:    $searched \leftarrow \{\}$
10:   **for** $node \in C_i$ **do**
11:      **if** $node \notin searched$ **then**
12:         $neighbor \leftarrow C_i \cup \{$neighbors of $node$ in $N\}\backslash searched$
13:         Remove all nodes with smaller node index than $node$ in $neighbor$
14:         $clique \leftarrow \boldsymbol{SearchTree}(\{node\}, neighbor, searched)$
15:         **if** size($clique$) $> 0$ **then**
16:            $CL \leftarrow CL \cup \{clique\}$
17:            $searched \leftarrow searched \cup clique$
18:         **end if**
19:      **end if**
20:   **end for**

21:     **return** $CL$

22:     **function** $\textit{SearchTree}$ $(subgraph, candidate, searched)$
23:         $clique \leftarrow subgraph$
24:         **while** size($candidate$) $> 0$ **do**
25:             $node \leftarrow$ first element from $candidate$
26:             Remove first element from $candidate$
27:             $neighbor \leftarrow C_i \cup \{$neighbors of $node\}\backslash searched$
28:             Remove all nodes with smaller index than $node$ in $neighbor$
29:             $newCandidate \leftarrow candidate$
30:             **for** $n \in neighbor$ **do**
31:                 **if** $n \notin newCandidate$ **then**
32:                     $newCandidate \leftarrow newCandidate \cup \{n\}$
33:                 **end if**
34:             **end for**
35:             $newSubgraph \leftarrow subgraph \cup \{newNode\}$
36:             **if** CID($N, newSubgraph$) $\geq CID\_threshold$ **then**
37:                 **if** size($newSubgraph$) $\geq Md$ **then**
38:                     **return** $newSubgraph$
39:                 **end if**
40:                 $result \leftarrow \textit{SearchTree}(newSubgraph, newCandidate, searched)$
41:                 **if** size($result$) $>$ size($clique$) **then**
42:                     $clique \leftarrow result$
43:                 **end if**
44:             **end if**
45:         **end while**
46:         **if** size($clique$) $\geq 3$ **then**
47:             **return** $clique$
48:         **end if**
49:         **return** {}
50:     **end function**

---

**Algorithm 2** Feature Migration

---

**Input:** Network $N$, cliques set $CL = \{clique_1, clique_2 ...\}$, transfer probability $Tp \in [0,1]$, population size $Ps \in \{1,2,3 ...\}$.

**Output:** Population set $Pop = \{solution_i | i = 1 \text{ to } Ps\}$

1:       $Pop = \{\}$
2:       Initialize *candidate* as a Hashmap
3:       **for** $clique \in CL$ **do**
4:          **for** $node \in clique$ **do**
5:            $neighbor \leftarrow clique \cup \{\text{neighbors of } node \text{ in } N\}\backslash\{node\}$
6:            $candidate.add(key = node, value = neighbor)$
7:          **end for**
8:       **end for**
9:       **for** $i \leftarrow 1 \text{ to } Ps$ **do**
10:      **for** $key\_n \in candidate$ **do**
11:         Sample **u** $\sim$ Uniform $(0, 1)$
12:         **if** $u \leq Tp$ **then**
13:            $candidateLabels \leftarrow candidate.value(key\_n)$
14:            **if** size($candidateLabels$) $> 0$ **then**
15:               Choose *label* from *candidateLabels* randomly
16:               $solution_i[key\_n] \leftarrow label$
17:            **end if**
18:         **end if**
19:      **end for**
20:      Run label propagation on $solution_i$ with 5 iterations
21:      $Pop \leftarrow Pop \cup solution_i$
22:     **end for**
23:     **return** $Pop$

---

**Algorithm 3** TMOGA

---

**Input:** Dynamic Networks $\{N^1, N^2, N^3 ... N^T\}$, population size $Ps \in \{1,2,3 ...\}$, number of generations $Gen \in \{1,2,3 ...\}$, threshold $CID\_threshold \in [0,1]$, max depth $Md \in \{1,2,3 ...\}$, transfer probability $Tp \in [0,1]$, crossover probability $Cp \in [0,1]$, mutation probability $Mp \in [0,1]$.

**Output:** Dynamic communities $Dc = \{CR^1, CR^2, CR^3 ... CR^T\}$

1:       $Dc \leftarrow \{\}$
2:       Initialize population $Pop^1$ of size $Ps$ by label propagation with 5 iterations
3:       **for** $k \leftarrow 1 \text{ to } Gen$ **do**
4:           Calculate Pareto rank for each individual

| 5: | Take selection operation to get 2 individuals from $Pop^1$ based on the weights as inverse Pareto rank |
|---|---|
| 6: | Take crossover operation on these 2 individuals to get 2 offspring |
| 7: | Take mutation operation on each genes of 2 offspring with probability $Mp$ |
| 8: | Add offspring to $Pop^1$. |
| 9: | Use elitist strategy to update population $Pop^1$. |
| 10: | **end for** |
| 11: | Select $CR^1$ with highest *Community Score* from Pareto solutions. |
| 12: | $Dc \leftarrow Dc \cup C^1$ |
| 13: | **for** $i \leftarrow 2$ to $T$ **do** |
| 14: | $Clique \leftarrow \{\}$ |
| 15: | **for** $c \in CR^{i-1}$ **do** |
| 16: | $CL \leftarrow \textbf{\textit{FeatureExtraction}}(N^{i-1}, c, CID\_threshold, Md)$ |
| 17: | $Clique \leftarrow Clique \cup CL$ |
| 18: | $Pop^i = \textbf{\textit{FeatureMigration}}(N^i, Clique, Tp, Ps)$ |
| 19: | **for** $k = 1$ to $Gen$ **do** |
| 20: | Calculate Pareto rank for each individual. |
| 21: | Take selection to get 2 individuals from $Pop^i$ based on the weights as inverse Pareto rank |
| 22: | Take crossover operation on these 2 individuals to get 2 offspring. |
| 23: | Take mutation operation on each genes of 2 offspring with probability $Mp$. |
| 24: | Add offspring to $Pop^i$. |
| 25: | Use elitist strategy to update population $Pop^i$. |
| 26: | **end for** |
| 27: | Select $CR^i$ with highest *Community Score* from Pareto solutions. |
| 28: | $Dc = Dc \cup CR^i$ |
| 29: | **end for** |
| 30: | **return** $Dc$ |

## 5.2. Implementation of Feature Migration

After cliques are obtained, we then propose a feature migration algorithm to add features to the next population. At the same time, in order to ensure the diversity of the population, we transfer the features only according to a certain probability, called *Transfer Probability* (*Tp*). High *Tp* may reduce the diversity in population, and 0.5 is set in our experiments in section 7. The process of the feature migration algorithm is to assign the community structure of each feature to the initial solution of each population randomly by *Tp*, thereby replacing the tags of these initial solutions. However, since the network structures can be changed either smoothly or dramatically between two continuous snapshots, we need to slightly adjust these features to be suitable to current network. Inspired by migration operator introduced by [7], we run label propagation algorithm [32] with only 5 iterations after cliques are assigned into current snapshot. In each iteration, the label of each node will be assigned as the label of most neighbors have. This additional step will typically enhance the quality of initial solutions. Comparable experiments results are provided in Section 7.6. The algorithm of feature migration is shown in Algorithm 2.

## 5.3. Multi-objective Evolutionary Algorithm

**Encoding** For community clustering problem, a number of different coding schemes have been proposed. Encoding can be broadly divided into two major categories: direct encoding and indirect encoding, as shown in figure 5. Direct encoding usually uses tag numbers to represent the number of societies to which each point belongs. The shortcoming is that direct encoding solution often loses some information in some operations. Indirect encoding, usually needs to be decoded into the corresponding encoding solution of the encoding scheme, such as locus-based adjacency encoding [33], whose decoding process can be completed in linear time [34]. In this encoding pattern, the label of node will be one of its neighbors in the same community. In decoding step, the locus-based solution will be naturally saved as the classic data structure disjoint-set. Thus the algorithms of disjoint-set can simply convert locus-based encoding to community structure. Also, in the feature migration algorithm, disjoint-set can help with label propagation algorithm. Because this kind of encoding can better retain the solution so that overall structure features will not be easily damaged, we use the locus-based adjacency encoding as the encoding scheme for the evolutionary algorithm.

**Initialization** To obtain initial solutions with greater quality, we use label propagation with only 5 iterations to generate initial states. Additionally, from the second snapshot on, the nodes in small cliques will be assigned the labels referring to feature transfer process.

**Selection** The original selection operation in NSGA-II is the traditional tournament method. However, the individuals with low fitness cannot have chance to crossover and mutate at all in such selection mechanism. To avoid this shortcoming, we use inverse Pareto rank as the weight for each individual and randomly select two according to their weights (see table 1).

**Crossover** The main method of uniform-crossing is to use both parents as prototypes to reproduce two children simultaneously (report in table 2). On the basis of the genotype of the parents, a binary mask vector as same length as a parent will be sampled from Binomial distribution. For each gene, the first child will get from first parent if the corresponding position in mask vector is 1, and get from second parent if 0. Then the second child will get gene from the parent left. In genetic algorithm, we used a *Crossover Probability* ($Cp$) to specify the parameter of Binomial distribution. When $Cp$ is close to 1 or 0, little crossover will happen; while $Cp$ is close to 0.5, crossover operation will be taken in most of genes.

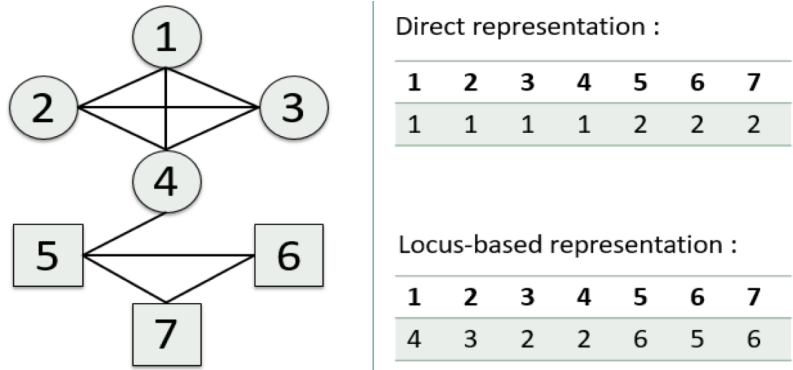

Fig. 5. Direct encoding and indirect encoding

**Mutation** This algorithm uses single point variation with a *Mutation Probability* (*Mp*). The process of single point variation is that after each generation of the descendants, each node in the descendant solution has a certain probability to randomly change the label of that point into the label of one of its neighboring nodes.

The overall procedure of our algorithm TMOGA is reported in Algorithm 3. Given dynamic networks and the required parameters for algorithm, TMOGA obtains the clustering result by running the genetic algorithm NSGA-II to optimize only one objective function, *Modularity*, when $t = 1$. Then from the second network on, the Feature Extraction and Feature Migration algorithms will start to work on the communities obtained at last snapshot, then generates initial population for current network. As repeated, NSGA-II will optimize both 2 objective functions *Modularity* and *NMI*. In each generation of genetic algorithm, selection operation, crossover operation and mutation operation will be taken sequentially. After that, a new pool of individuals will be created by parents and offspring. Based on elitist strategy of NSGA-II, only the solutions with high fitness will retain. When the termination condition of genetic algorithm is satisfied, we will obtain a set of Pareto solutions, then the solution with highest *Community Score* will be selected as the final solution for current network, as discussed before in section 3.6.

| solutions | 1 | 2 | 3 | 4 | 5 | 6 | 7 |
|---|---|---|---|---|---|---|---|
| **Pareto ranks** | 1 | 3 | 2 | 1 | 2 | 4 | 5 |
| **Unnormalized weight** | 1 | 1/3 | 1/2 | 1 | 1/2 | 1/4 | 1/5 |

Table 1. Example of Selection Weights

| Parent 1 | 3 | 5 | 1 | 1 | 3 | 4 | 6 |
|---|---|---|---|---|---|---|---|
| Parent 2 | 1 | 3 | 2 | 6 | 2 | 4 | 5 |
| Mask | 0 | 1 | 1 | 1 | 0 | 0 | 1 |
| Offspring 1 | 1 | 5 | 1 | 1 | 2 | 4 | 6 |
| Offspring 2 | 3 | 3 | 2 | 6 | 3 | 4 | 5 |

Table 2. Example of Uniform Crossover

## 6. Information-Theoretical Analysis

This section introduces a new theoretical framework to analyze dynamic community discovery problem. The traditional theoretical analysis heavily depends on network science. Although it has been developed for a long time, it relies on specific evaluation criteria and network data, and is difficult to analyze this problem uniformly. Thus, we seek for another new framework from informational perspective. Based on information theory, a variational objective function proposed in [11] was utilized to unify different criteria for evaluation of community results, e.g. *Modularity* [22], *Community Score* [23], etc. From the informational view, we also prove that feature transfer can dramatically improve the clustering results, both on optimization of snapshot cost and temporal cost.

### 6.1. Prerequisite: Information Bottleneck Theory

*Information Bottleneck* theory is a probabilistic framework to describe the process of information compression, built on variational principle. It is formally proposed in literature [11], and widely utilized to analyze algorithms of machine learning [35, 36]. This section will introduce some basic background of this theory and use it to prove some mathematical theorems in next subsection.

Let $X$ be a discrete random variable with distribution $p$, whose value is in $\{x_1, x_2, \ldots, x_n\}$. According to information entropy introduced by Shannon [37]. The information provided by a random variable can be measured as:

$$H(X) = -\sum_{i=1}^{n} p(x_i) * log\ (p(x_i)) \qquad (6)$$

In some practical scenarios, signals (denoted by random variable $X$) are too vague and the information carried should be compressed to denoise, e.g. eliminating the noise in video. In this traditional information compression problem, we need to find another discrete random variable $\tilde{X}$, to denote the information left in $X$ after compression. For each value in $X$, we can seek a stochastic mapping to an element in $\tilde{X}$, characterized by conditional probability

$p(\tilde{x}|x)$. For measurement of the information uncompressed, *mutual information* is proposed in [38]:

$$I(X; \tilde{X}) = \sum_{x \in X} \sum_{\tilde{x} \in \tilde{X}} p(x, \tilde{x}) log \frac{p(x, \tilde{x})}{p(x)p(\tilde{x})} \tag{7}$$

*Mutual information* quantifies the correlated information between $X$ and $\tilde{X}$. Less *mutual information* indicates more information has been compressed. Obviously, we can find $I(X; X) = H(X)$ and $I(X; \tilde{X}) = I(\tilde{X}; X)$.

It is the objective of information compression to eliminate as much redundant information as possible while keeping "relevant" information from being lost, e.g. eliminating more noise while keeping more voice. Obviously, this is a trade-off between compressing more information and preserving "meaningful" information. From what has been discussed in literature [11], let $Y$ denote a random variable which is not independent to $X$, where $I(X; Y)$ denotes the "relevant" information stored in X and is constant after giving $X$ and $Y$. Similarly, we can use $I(\tilde{X}; Y)$ to quantify "meaningful" information after compression, which should be maximized during compression process. It is easy to derive the inequality $I(\tilde{X}; Y) \leq I(X; Y)$. Intuitively, in this compression model, a part of the information regarding $Y$ provided by $X$ passes through a "bottleneck" (see Fig.6). The width of "bottleneck" should be controlled to block a large amount of irrelevant information and to maximize the amount of meaningful information passed. Consequently, to model this whole process, an objective function has been proposed in [11]:

$$\min_{p(\tilde{x}|x)} \mathcal{L}(\beta) = \min_{p(\tilde{x}|x)} \left[ I(X; \tilde{X}) - \beta I(\tilde{X}; Y) \right], \beta \geq 0 \tag{8}$$

where $\beta$ is the Lagrange multiplier to balance the trade-off between these 2 objectives and controls the width of "bottleneck". In [11], authors applied the variational principle and the Markovian property of information compression, to find the optimal solution $\hat{p}(\tilde{x}|x)$ of function (8):

$$\hat{p}(\tilde{x}|x) = \frac{p(\tilde{x}) \exp\left(-\beta D_{kl}[p(y|x) \,||\, p(y|\tilde{x})]\right)}{Z(x, \beta)} \tag{9}$$

Where $D_{kl}(p||q)$ is the well-known KL divergence for distributions $p$ and $q$, defined in literature [39], and $Z(x, \beta) = \sum_{\tilde{x}} \{p(\tilde{x})exp(-\beta D_{kl}[p(y|x) \,||\, p(y|\tilde{x})])\}$ is the normalization function for the numerator in (9).

*6.2. Probabilistic Framework to Analyze Community Detection*

Inspired by *Information Bottleneck* method, we use probabilistic framework to redefine the dynamic community detection problem.

Suppose at snapshot t, network $N^t$ has exact $n^t$ nodes. Let $X^t$ be a discrete uniform random variable with distribution $p$, whose value varies in $\{1, 2, \ldots, n^t\}$. In some sense, $X^t$ can be a representation for network $N^t$, and $p(X^t = i) = \frac{1}{n^t}$. Suppose that network $N^t$ is divided into $k$ disjoint communities $CR^t = \{C_1^t, C_2^t, \ldots, C_k^t\}$. With the same reason, we define another discrete random variable $\tilde{X}^t$ to represent communities $CR^t$, whose value varies in $\{1, 2, \ldots, k\}$. Naturally, we can give the conditional probability $p(\tilde{x}^t | x^t)$, which maps the nodes in $X^t$ to the communities associated with them in network $N^t$. In other word, $p(\tilde{X}^t = i | X^t = j)$ accounts for the probability that the *j*-th node is mapped to *i*-th community at the *t*-th snapshot. Although this probabilistic framework introduces uncertainty in the deterministic community detection problem, we can simply set $p(\tilde{X}^t = i | X^t = j) = 1$ if *j*-th node belongs to *i*-th community on ground truth and $p(\tilde{X}^t = k | X^t = j) = 0$ if not. Therefore, the problem of community detection on dynamic network can be redefined as:

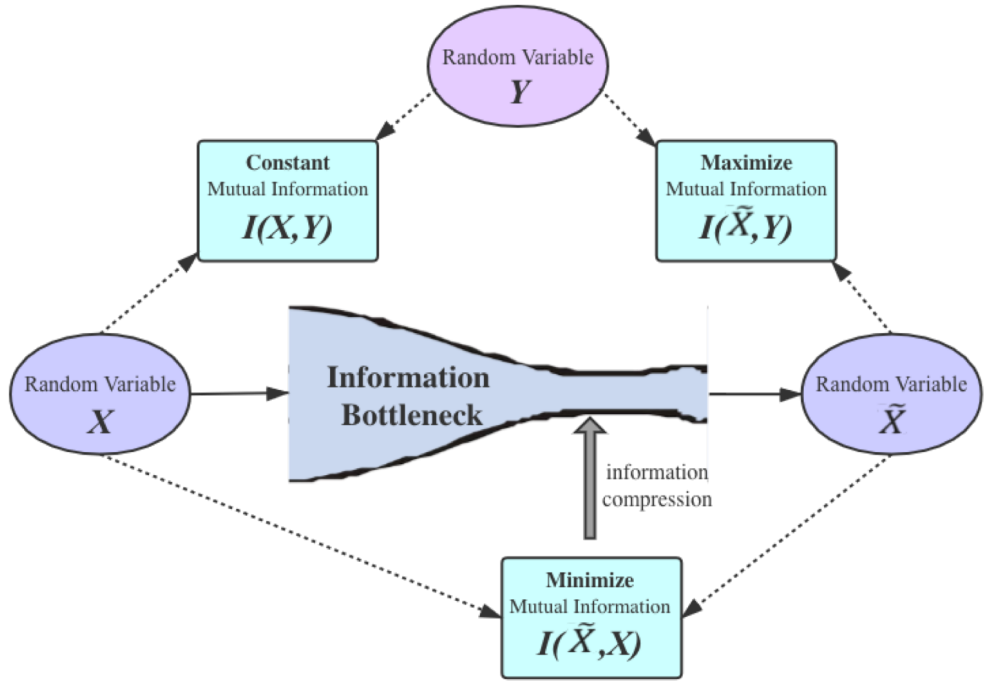

Fig. 6. Paradigm for *Information Bottleneck*

$$\underset{p(\tilde{x}^t|x^t)}{\text{argmin}} \begin{cases} -SC(\bar{X}^t) \\ -NMI(\bar{X}^t; \bar{X}^{t-1}) \end{cases} \tag{10}$$

Here $SC$ is snapshot cost, which can be any function to evaluate community result of single network, such as *Modularity*, *Community Score*, etc. It is worthwhile to note, although function $SC(N^t)$ is defined on a network, we can simply find another function $\widetilde{SC}(X^t)$ that maps variable $X^t$ to the quantity of $SC(N^t)$, i.e $\widetilde{SC}(X^t) = SC(N^t)$, where $X^t$ is the representation of network $N^t$. To simplify the notation, we will use $SC(X^t)$ to replace $\widetilde{SC}(X^t)$, and *NMI* defined in (2) as well.

After the redefinition into probabilistic problem, the *Information Bottleneck* theory can be applied to analyze the problem of dynamic community detection. We will then prove two theorems (we drop the superscript $t$ unless it is needed) to discuss the equivalence of the optimization of snapshot cost $-SC(\bar{X}^t)$ and objective function (8).

**Theorem 1.** For any partition $p(\tilde{x}|x)$, the corresponding communities variable $\tilde{X}$ has smaller or equal information entropy than $X$, which gives following inequality:

$$H(\tilde{X}) \leq H(X), \text{as } X \sim \text{ discrete uniform distrinution} \tag{11}$$

The equality will hold if $\tilde{X}$ has same distribution as $X$.

**Proof of Theorem 1.** From what has been discussed in literature [40], uniform distribution with no constraint has maximum entropy among all discrete variables. Since $X$ has discrete uniform distribution, so any discrete variable $\tilde{X}$ follows formula (11), which completes the proof.

This theorem shows any partition of network will lose information entropy unless the partition does nothing on network, which assigns every node to the community consists of only itself. It indicates clustering is a process of information compression which refines information from node level to community level. Therefore, *Information Bottleneck* method can be applied explicitly to analyze community detection problem in this probabilistic framework.

**Theorem 2.** At timestamp $t$, let $\hat{p}(\tilde{x}|x)$ denote the solution for (10). Then, there exists at least one specific $\hat{\beta} \geq 0$, such that $\hat{p}(\tilde{x}|x)$ is the solution for $\underset{p(\tilde{x}|x)}{\min} \mathcal{L}(\hat{\beta})$ defined in (9): $\hat{p}(\tilde{x}|x) = \underset{p(\tilde{x}|x)}{\text{argmin}} \mathcal{L}(\hat{\beta})$.

**Proof of Theorem 2.** The theorem is obvious, since $\hat{p}(\tilde{x}|x)$ is a function of $\beta$ in (9), where this function is a surjective mapping evidently. So there exists at least one $\hat{\beta}$, which is mapped to $\hat{p}(\tilde{x}|x)$. Thus we complete the proof.

This theorem is crucial for our informational framework. It reveals the fact that the solution for any *SC* function is also the solution for information bottleneck with some specific $\beta$. In other word, the optimization of any *SC* function is equivalent to the optimization of *Information Bottleneck* model. Therefore, all *SC* functions can be unified into a probabilistic framework and analyzed uniformly. With this significant connection, we can directly analyze *Information Bottleneck* instead of any specific *SC* function.

To illustrate this theorem intuitively, we reformulate *Information Bottleneck* according to community detection problem. Since there is no other variable $Y$ providing information about $X$ except $\tilde{X}$, we use $\tilde{X}$ as $Y$ in function (8):

$$\min_{p(\tilde{x}|x)} \mathcal{L}(\beta) = \min_{p(\tilde{x}|x)} \left[ I(X; \tilde{X}) - \beta I(\tilde{X}; \tilde{X}) \right] = \min_{p(\tilde{x}|x)} \left[ I(X; \tilde{X}) - \beta H(\tilde{X}) \right] \tag{12}$$

From the property discussed in section 6.1, function (12) indicates we should compress more information of $X$ while maximize the information of $\tilde{X}$. Therefore, if $\beta = 0$, the problem becomes $I(X; \tilde{X}) = 0$, to compress all information in X. So the optimal partition is the most sketchy solution, $H(\tilde{X}) = 0$, and the whole network is a community. This solution is that of Min-entropy problem without any constrains. On the contrary, as $\beta \to \infty$, function (12) is equivalent to maximizing $H(\tilde{X})$. In this case, $\tilde{X}$ follows a discrete uniform distribution that $\tilde{X} = X$ and every node belongs to the community of itself. This scenario corresponds to solution of Max-entropy problem.

Here we have proposed the new framework to describe dynamic community detection successfully. To avoid analyzing any specific *SC* function directly, we first show the connection between them and *Information Bottleneck* and then we can analyze *Information Bottleneck* function (8) instead. There are mainly three advantages: 1. Avoid analyzing different criteria respectively; 2. *Information Bottleneck* function (8) is a differentiable and continuous function in regard to $p(\tilde{x}|x)$ while most of the *SC* functions are discrete; 3. The best solution of the *SC* functions significantly depends on the network data while the solution of *Information Bottleneck* can be directly obtained, which can reduce the complexity to analyze.

### 6.3. Proof of Rationality of Feature Transfer

After we proposed the informational framework to analyze community detection problem, we then utilize it to prove the fact that feature transfer can enhance the result of dynamic community detection significantly.

**Definition 1: Small Cliques(Features).** We assume *Small Cliques* are unchanged through the iterations of NSGA-II, indicating *Small Cliques* are parts of the final solution for both $\tilde{X}^t$ and $\tilde{X}^{t-1}$. Let random variable $Z^t$ consist of those subgraphs remains stable in $\tilde{X}^t$ and $\tilde{X}^{t-1}$, i.e. $p(z^t|X^t = j) = p(\tilde{x}^t|X^t = j) = p(\tilde{x}^{t-1}|\tilde{X}^{t-1} = j)$ for some nodes $j$ in *Small Cliques* set in networks $X^t$ and $\tilde{X}^{t-1}$.

**Theorem 3.** At any snapshot $t$, for any given $\beta \geq 0$, feature transfer has higher compression efficiency which indicates the following inequality (we drop the superscript $t$ unless it is needed):

$$\left[I(X, Z; \tilde{X}) - \beta I(\tilde{X}; \tilde{X})\right] \leq \left[I(X; \tilde{X}) - \beta I(\tilde{X}; \tilde{X})\right] \tag{13}$$

**Proof of Theorem 3.** Since $\beta \geq 0$, inequality (13) can be reformulated as:

$$I(X; \tilde{X}) - I(X, Z; \tilde{X}) \geq 0 \tag{14}$$

According to function (7), we have the following equation:

$$
\begin{aligned}
I(X; \tilde{X}) &- I(X, Z; \tilde{X}) \\
&= \sum_{\tilde{x} \in \tilde{X}} \sum_{x \in X} p(x, \tilde{x}) log \frac{p(x, \tilde{x})}{p(x)p(\tilde{x})} \\
&\quad - \sum_{\tilde{x} \in \tilde{X}} \sum_{x \in X} \sum_{z \in Z} p(x, z, \tilde{x}) log \frac{p(x, z, \tilde{x})}{p(x, z)p(\tilde{x})} \\
&= \sum_{\tilde{x} \in \tilde{X}} \sum_{x \in X} \left[ p(x, \tilde{x}) log \frac{p(x, \tilde{x})}{p(x)p(\tilde{x})} - \sum_{z \in Z} p(x, z, \tilde{x}) log \frac{p(x, z, \tilde{x})}{p(x, z)p(\tilde{x})} \right]
\end{aligned}
\tag{15}
$$

Since $p(x, z, \tilde{x}) = p(x, \tilde{x})$ when $p(z|x) = p(\tilde{x}|x)$ and $p(x, z, \tilde{x}) = 0$ elsewhere. Therefore, for a fixed $\tilde{x}$ and $x$, $\sum_{z \in Z} p(x, z, \tilde{x}) log \frac{p(x, z, \tilde{x})}{p(x, z)p(\tilde{x})} = p(x, \tilde{x}) log \frac{p(x, \tilde{x})}{p(x, z)p(\tilde{x})}$ holds and equation (15) can be written as:

$$I(X; \tilde{X}) - I(X, Z; \tilde{X}) = \sum_{\tilde{x} \in \tilde{X}} \sum_{x \in X} \left[ p(x, \tilde{x}) log \frac{p(x, \tilde{x})}{p(x)} \right] = D_{KL}(p(x, \tilde{x})||p(x)) \geq 0 \tag{16}$$

Where $D_{KL}(p||q)$ denotes the KL divergence which is non-negative everywhere. Thus, we complete the proof that feature transfer can enhance the optimization of *Information Bottleneck* function and *SC* as well.

**Theorem 4.** Feature transfer can also optimize more effectively on another objective function $NMI(\tilde{X}^t; \tilde{X}^{t-1})$:

$$NMI(\tilde{X}^t, Z^t; \tilde{X}^{t-1}) \geq NMI(\tilde{X}^t; \tilde{X}^{t-1}) \tag{17}$$

**Proof of Theorem 4.** The expression of $NMI(A; B)$ has been shown as function (2). As a matter of fact, function (2) is the discrete form. Generally, the probabilistic expression of $NMI(A; B)$ is as follow:

$$NMI(A; B) = \frac{2I(A; B)}{H(A) + H(B)} \qquad (18)$$

Here we use the same probabilistic framework as above to prove this theorem and reformulate inequality (17) as:

$$\frac{2I(\tilde{X}^t, Z^t; \tilde{X}^{t-1})}{H(\tilde{X}^t, Z^t) + H(\tilde{X}^{t-1})} \geq \frac{2I(\tilde{X}^t; \tilde{X}^{t-1})}{H(\tilde{X}^t) + H(\tilde{X}^{t-1})} \qquad (19)$$

Since $I(A; B) = H(A) + H(B) - H(A, B)$ where $H(A, B)$ is the joint entropy [38] inequality (19) takes the form:

$$\frac{H(\tilde{X}^t, Z^t, \tilde{X}^{t-1})}{H(\tilde{X}^t, Z^t) + H(\tilde{X}^{t-1})} \leq \frac{H(\tilde{X}^t, \tilde{X}^{t-1})}{H(\tilde{X}^t) + H(\tilde{X}^{t-1})} \qquad (20)$$

Where $H(\tilde{X}^t, Z^t, \tilde{X}^{t-1}) = -\sum_{\tilde{x}^t}\sum_{\tilde{x}^{t-1}}\sum_{z^t} p(\tilde{x}^t, z^t, \tilde{x}^{t-1}) \log p(\tilde{x}^t, z^t, \tilde{x}^{t-1})$. Consequently, inequality (20) is what we want to prove. Apparently, $p(\tilde{x}^t, z^t, \tilde{x}^{t-1}) \log p(\tilde{x}^t, z^t, \tilde{x}^{t-1}) = p(\tilde{x}^t, \tilde{x}^{t-1}) \log p(\tilde{x}^t, \tilde{x}^{t-1})$ when $p(z^t|x^t) = p(\tilde{x}^t|x^t)$ and $p(\tilde{x}^t, z^t, \tilde{x}^{t-1}) \log p(\tilde{x}^t, z^t, \tilde{x}^{t-1}) = 0$ otherwise. Therefore, for a fixed $\tilde{x}^t$ and $\tilde{x}^{t-1}$, $\sum_{z^t} p(\tilde{x}^t, \tilde{x}^{t-1}) \log p(\tilde{x}^t, z^t, \tilde{x}^{t-1}) = p(\tilde{x}^t, \tilde{x}^{t-1}) \log p(\tilde{x}^t, \tilde{x}^{t-1})$ holds. Formally, we have this equation:

$$H(\tilde{X}^t, Z^t, \tilde{X}^{t-1}) = -\sum_{\tilde{x}^t}\sum_{\tilde{x}^{t-1}} p(\tilde{x}^t, \tilde{x}^{t-1}) \log p(\tilde{x}^t, \tilde{x}^{t-1}) = H(\tilde{X}^t, \tilde{X}^{t-1}) \qquad (21)$$

Owing to $H(A, B) = H(A) + H(B|A)$ where $H(B|A) \geq 0$ discussed in literature [38], thus we have $H(\tilde{X}^t, Z^t) \geq H(\tilde{X}^t)$. Evidently, combining with equation (21), we find:

$$\frac{H(\tilde{X}^t, Z^t, \tilde{X}^{t-1})}{H(\tilde{X}^t, Z^t) + H(\tilde{X}^{t-1})} \leq \frac{H(\tilde{X}^t, Z^t, \tilde{X}^{t-1})}{H(\tilde{X}^t) + H(\tilde{X}^{t-1})} = \frac{H(\tilde{X}^t, \tilde{X}^{t-1})}{H(\tilde{X}^t) + H(\tilde{X}^{t-1})} \qquad (22)$$

The inequality (22) is what we want to prove for inequality (20) which show transfer learning has a higher efficiency to optimize the objective function of *NMI*. This completes the proof.

To conclude, in this section, we prove 2 theorems built on informational theoretical framework, where *Theorem 3* proves the improvement from feature transfer on the optimization of *Information Bottleneck* function and thus *SC* function; *Theorem 4* proves that feature transfer can accelerate optimizations of *NMI* as well. Totally, feature transfer can improve the optimization of objective function (10) simultaneously.

## 7. Experiment

This chapter proves the practical effect of TMOGA through the experiment. The control experiments include the DYNMOGA [3] algorithm, proposed by Folino in 2014, which is based on NSGA-II framework; DECS [7] proposed in 2020, is another algorithm based on NSGA-II framework, which encodes solution directly in adjacency matrix to identify edges inside/outside communities to separate clusters and it uses an edge migration operation to improve results; DYNMODPSO [8] is a heuristic algorithm based on *temporal smoothness* framework, which initializes population by random walk and applies particle swarm algorithm to enhance the optimization of snapshot cost, it also utilizes a special correction mechanism to get solutions with high quality at $t = 1$ and $t = 2$; FacetNet algorithm [13] proposed a bunch-based mathematical approach with very fast speed, which supposes communities as hidden nodes. The experiments are conducted on a MacBook Pro with 8G memory and 2.3 GHz Intel Core i5 CPU. All algorithms are implemented in Python.

### 7.1. Parameters setting

As introduced above, in our algorithm TMOGA, we have 7 parameters to control, including Population, Generations, *CID* threshold, Max Depth of search tree (*Md*), Transfer Probability (*Tp*), and Crossover Probability (*Cp*), Mutation Probability (*Mp*) of NSGA-II. To conclude, after careful parameter selection, we directly employed parameters in table 3 in the following experiments.

| Parameters | Value |
|:---|:---:|
| Population size | 200 |
| Generations | 100 |
| *CID* threshold | 0.8 |
| Max depth of search tree(*Md*) | 5 |
| Transfer probability (*Tp*) | 0.5 |
| Crossover probability (*Cp*) | 0.8 |
| Mutation probability (*Mp*) | 0.2 |

Table 3. Parameters of Experiments

Since compared algorithms except FacetNet are all based on heuristic algorithm, we set the population size as 200 and generation as 100 for equality. For DYNMOGA, *Cp*, *Mp* are set the same as TMOGA. We set the parameter α balancing *SC* and *TC* as 0.5 for FacetNet and specify the number of communities by running semi-synchronous label propagation method [41]. Finally for DECS and DYNMODPSO, other parameters are set the same as them in papers [7] and [8], respectively.

## 7.2. Evaluation Criteria

We use multiple criteria to evaluate the experimental results. For data with ground truth, the evaluation criteria are *NMI* between the solution and the real result from data. When *NMI* is close to 1, it shows that the solution obtained is equivalent to the real solution, that is to say, the perfect one. It is worth noting that *NMI* here is used to evaluate the final solution obtained from TMOGA algorithm, not for the evaluation of the community partitions between two continuous snapshots.

For data without labels, we simply evaluate the best solution for different algorithms by *Modularity*. The higher *Modularity* indicates the better algorithm on this dataset.

In addition, the running time of each algorithm will also be recorded. Although the time cannot be recorded rather accurate, the magnitude can reflect the relative speed of algorithms. We additionally record the total time of feature transfer, and the results show the time of feature transfer can be ignored compared with the running time of NSGA-II.

## 7.3. Benchmark Dataset

The first dataset of experiments came from literature [13], which is based on the generating method of the classic test set proposed by Girvan and Newman [1]. A parameter Z controls the number of edges connected to each community. We apply *NMI* to evaluate the results.

**SYNFIX** The dataset is set with the fixed number of communities. The test set contains 4 communities, 32 nodes per community, and 128 nodes in all. At each point in time, 3 nodes in each community are randomly selected, and they are randomly assigned into another community to form a new network.

**SYNVAR** This dataset changes the fixed number of communities through time. The entire network in the data set contains 256 nodes, 4 communities, and 64 nodes per community. There are 10 snapshots in total. This dataset randomly selected 8 nodes from each community in the initial moment to form a new community. Repeat this operation 4 times, and get the first 5 snapshots including the initial snapshot. On the fifth snapshot, there are 8 communities in total, and 32 nodes in each community. Then at snapshot 6, it keeps same number of community as snapshot 5. From the beginning of the 7th time we move the node back to the original community to generate other 5 snapshots.

| SYNFIX Z = 3 | | | | | | | | | | | |
|---|---|---|---|---|---|---|---|---|---|---|---|
| Algorithm | 1 | 2 | 3 | 4 | 5 | 6 | 7 | 8 | 9 | 10 | Time(s) |
| TMOGA | **1** | **1** | **1** | **1** | **1** | **1** | **1** | **1** | **1** | **1** | 1836(25.3) |
| DECS | 1 | 1 | 1 | 1 | 1 | 1 | 1 | 1 | 1 | 1 | 2245 |
| FacetNet | 0.857 | 0.860 | 0.860 | 0.860 | 0.860 | 0.858 | 0.860 | 0.831 | 0.857 | 0.860 | 13 |
| DYNMODPSO | 0.979 | 0.979 | 0.979 | 0.979 | 0.979 | 0.979 | 0.979 | 0.979 | 0.979 | 0.979 | 2499 |
| DYNMOGA | 0.975 | 1 | 1 | 1 | 1 | 1 | 1 | 1 | 1 | 1 | 1789 |
| SYNFIX Z = 6 | | | | | | | | | | | |
| Algorithm | 1 | 2 | 3 | 4 | 5 | 6 | 7 | 8 | 9 | 10 | Time(s) |
| TMOGA | **1** | **1** | **1** | **1** | **1** | **1** | **1** | **1** | **1** | **1** | 1831(24.8) |
| DECS | 1 | 1 | 1 | 1 | 0.957 | 1 | 1 | 1 | 1 | 1 | 2296 |
| FacetNet | 0.667 | 0.667 | 0.668 | 0.669 | 0.635 | 0.534 | 0.667 | 0.634 | 0.667 | 0.668 | 13 |
| DYNMODPSO | 0.988 | 1 | 0.852 | 0.649 | 0.922 | 0.743 | 0.661 | 0.810 | 0.929 | 0.656 | 2444 |
| DYNMOGA | 0.886 | 0.962 | 0.950 | 0.950 | 0.929 | 0.975 | 0.942 | 0.950 | 0.975 | 1 | 1896 |

Table 4. Clustering results and running time for SYNFIX. Time in parentheses means the running time of feature transfer.

| SYNVAR Z = 3 | | | | | | | | | | | |
|---|---|---|---|---|---|---|---|---|---|---|---|
| Algorithm | 1 | 2 | 3 | 4 | 5 | 6 | 7 | 8 | 9 | 10 | Time(s) |
| TMOGA | **1** | **1** | **1** | **1** | **1** | **1** | **1** | **1** | **1** | **1** | 3715(69.9) |
| DECS | 1 | 1 | 1 | 1 | 0.945 | 0.945 | 1 | 1 | 1 | 1 | 7637 |
| FacetNet | 1 | 0.815 | 0.759 | 0.702 | 0.741 | 0.8 | 0.824 | 0.866 | 0.923 | 1 | 69 |
| DYNMODPSO | 1 | 0.989 | 1 | 0.975 | 0.975 | 1 | 0.995 | 0.966 | 0.889 | 0.973 | 4685 |
| DYNMOGA | 0.934 | 1 | 1 | 1 | 1 | 1 | 1 | 0.99 | 1 | 1 | 3577 |
| SYNVAR Z = 6 | | | | | | | | | | | |
| Algorithm | 1 | 2 | 3 | 4 | 5 | 6 | 7 | 8 | 9 | 10 | Time(s) |
| TMOGA | **0.985** | **0.988** | **1** | **1** | **1** | **1** | **1** | **1** | **1** | **1** | 3683(71.9) |
| DECS | 0.901 | 0.916 | 1 | 1 | 1 | 0.988 | 0.962 | 0.921 | 0.909 | 0.882 | 7552 |
| FacetNet | 0.667 | 0.532 | 0.421 | 0.346 | 0.375 | 0.416 | 0.467 | 0.546 | 0.602 | 0.647 | 24 |
| DYNMODPSO | 0.896 | 0.937 | 0.755 | 0.892 | 0.969 | 0.962 | 0.864 | 0.861 | 0.659 | 0.602 | 4694 |
| DYNMOGA | 0.459 | 0.58 | 0.744 | 0.832 | 0.813 | 0.921 | 0.871 | 0.842 | 0.878 | 0.586 | 3638 |

Table 5. Clustering results and running time for SYNVAR. Time in parentheses means the running time of feature transfer.

**Result** As shown in table 4, the experiment shows that in the SYNFIX data, in both the solutions when Z = 3 and Z = 6, our algorithm TMOGA can obtains the optimal solutions in every snapshot with desirable running time. DECS, although the running time is unacceptable, the accuracy is remarkable that it obtains optimal solutions in almost every step. FacetNet has ignorable running time while its accuracy is the lowest. It is worthwhile to note for DYNMODPSO: it can obtain the great partitions in some snapshots while it has the highest variance that is unstable for other snapshots. DYNMOGA obtains solutions with a little lower accuracy as expected.

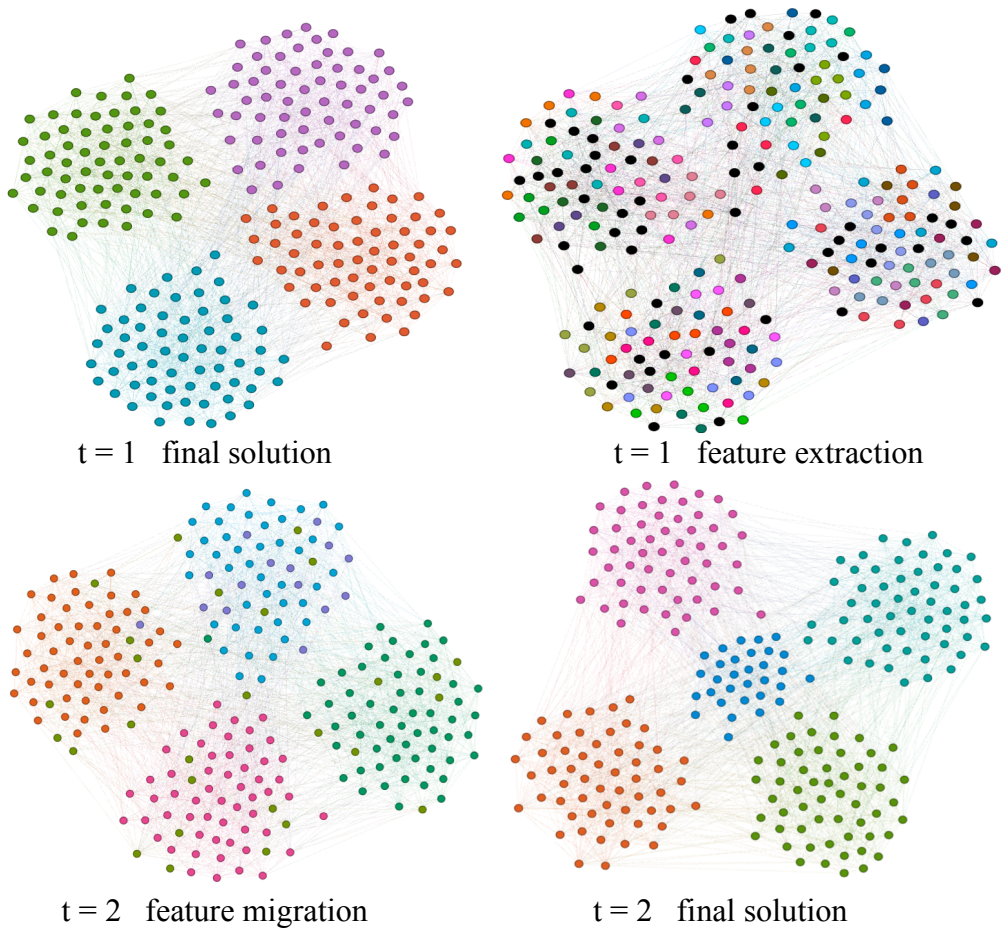

| t = 1   final solution | t = 1   feature extraction |

| t = 2   feature migration | t = 2   final solution |

Fig. 7. SYNVAR Z = 6, feature transfer for t = 1, 2. Black nodes belongs to no clique. Layout is adjusted for communities' visualization.

For the SYNVAR problem (see table 5), the TMOGA algorithm is still able to obtain the same solutions as real label except the first 2 snapshots when Z = 6. DECS has suboptimal performance that can identify the real community structures at some snapshots; but due to the increasing number of nodes, the encoding scheme of DECS needs to consume much more time. FacetNet, with the same performance in SYNFIX, has the worst accuracy, especially when the number of community changes. On the contrary, DYNMODPSO has good performance at first two snapshots due to its special correction mechanism but its accuracy decays at the subsequential networks. DYNMOGA, due to its naïve mechanism, has lower accuracy.

Figure 7 shows the visualization of feature transfer obtained from TMOGA for SYNVAR problem Z = 6. After feature extraction at t = 1, only few nodes belong to no clique, and most of subgraphs are structured. This phenomenon proves the prerequisite of feature transfer. Then after feature migration at t = 2, most of community structures are identified. It improves the quality of initial solutions significantly and simplifies the whole optimization process even if search space is complex. Consequently, feature transfer proves the notable performance of TMOGA.

The experiment shows that our TMOGA algorithm has undefeated convergence on small scale problems, and almost always gets the optimal solution with little time.

### 7.4. Large-scale Problem

This part tests the performance of the algorithm on a large scale dataset and *NMI* are used for evaluation. We use tool proposed in [42] to generate 4 types of dynamic networks[1]. In each dynamic network, there are 5 time steps and each snapshot network consists of 1000 nodes and about 33 communities. At each time step, 20% nodes will be assigned randomly to other community. After that, 4 distinct types of evolution events will occur in 4 dynamic networks respectively:

*Birth and Death Model*: 10% of nodes are taken from existing communities, resulting in new communities, while 10% of the community is removed.

*Expansion and Contraction Model:* At each moment, 10% of the community in the original communities are expanded by 25% of the original scale; 10 % are reduced by 25% of the original scale.

*Intermittent Communities Model:* At each snapshot, 10% of communities is hidden and appears in next time step.

*Merging and Splitting Model:* At each moment, 10% of the communities split, and 10% of the communities are selected with pair merging.

**Result** According to the experimental results, in general, our TMOGA algorithm is superior to the other four algorithms, in both *NMI* and running time. In table 6, the effect of TMOGA

---

[1] Tool can be found at http://mlg.ucd.ie/dynamic/ . Parameters for generating data are attached in Appendix. A

on the first snapshot is not remarkable, however, it achieves the best performance in the subsequential networks with the improvement by feature transfer. Also, the performance of TMOGA remains significantly stable, even if about 20% membership is changed at each step. For DECS, it obtains the notable partitions in most of the snapshots. However, it has two shortcomings: the longest computation time and decaying accuracy. FacetNet, unlike in other experiments, achieves the optimal results at the first snapshots in all dataset, and the suboptimal solutions at other networks. The reason why FacetNet can defeat other algorithms except TMOGA is that, small error caused by semi-synchronous label propagation method may lead to huge difference in small dataset; in large dataset, however, with the nearly accurate number of communities identified, the accuracy of FacetNet is enhanced dramatically. On the other hand, FacetNet is not stable enough, and consumes much more time with the increasing number of communities. DYNMODPSO has the stable performance in the first two networks in all dataset due to special correction mechanism, but does not perform well in other networks. DYNMOGA, unlike in other datasets, obtains the worst partitions at all steps in all datasets. It shows DYNMOGA mechanism has poor upper limit in large dataset.

Consequently, it is shown that the feature transfer of our algorithm is most suitable for the large datasets. TMOGA can achieve the best and stable results in all large datasets, with nearly little running time.

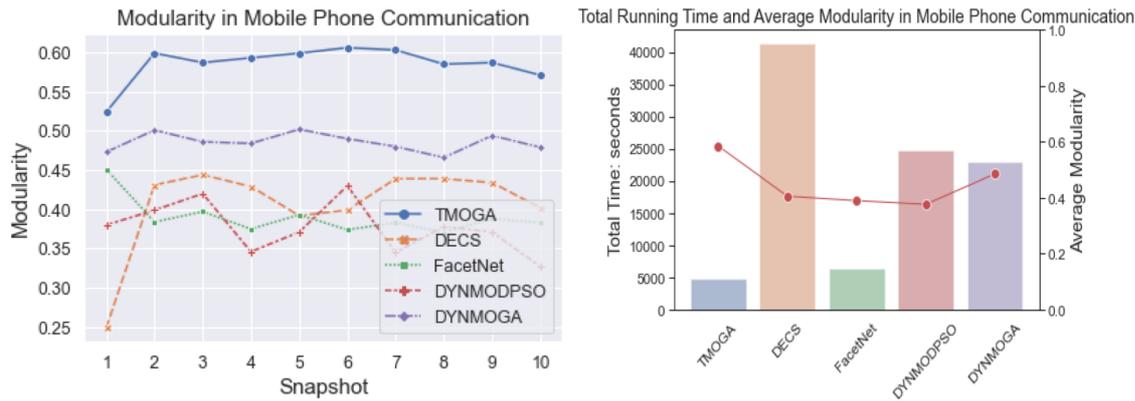

Fig. 8. Clustering results and running time for Mobile Phone Communication
In right panel, bars are running time and line is average *Modularity* for each algorithm

| Birth and Death | | | | | | |
|---|---|---|---|---|---|---|
| Algorithm | 1 | 2 | 3 | 4 | 5 | Time(s) |
| TMOGA | 0.919 | **0.929** | **0.936** | **0.936** | **0.945** | 6103(52.3) |
| DECS | 0.884 | 0.888 | 0.844 | 0.831 | 0.802 | 126695 |
| FacetNet | **0.967** | 0.927 | 0.754 | 0.925 | 0.898 | 15572 |
| DYNMODPSO | 0.832 | 0.855 | 0.793 | 0.809 | 0.811 | 34389 |
| DYNMOGA | 0.757 | 0.770 | 0.769 | 0.754 | 0.769 | 15886 |
| Expansion and Contraction | | | | | | |
| Algorithm | 1 | 2 | 3 | 4 | 5 | Time(s) |
| TMOGA | 0.927 | 0.933 | **0.934** | **0.944** | **0.971** | 6266(53.2) |
| DECS | 0.884 | 0.899 | 0.888 | 0.912 | 0.877 | 129454 |
| FacetNet | **0.969** | **0.972** | 0.688 | 0.858 | 0.836 | 19256 |
| DYNMODPSO | 0.811 | 0.837 | 0.788 | 0.779 | 0.802 | 36564 |
| DYNMOGA | 0.752 | 0.762 | 0.735 | 0.757 | 0.761 | 16998 |
| Intermittent Communities | | | | | | |
| Algorithm | 1 | 2 | 3 | 4 | 5 | Time(s) |
| TMOGA | 0.914 | **0.925** | **0.932** | **0.948** | **0.946** | 5858(51.3) |
| DECS | 0.906 | 0.842 | 0.828 | 0.794 | 0.834 | 128708 |
| FacetNet | **0.943** | 0.918 | 0.789 | 0.833 | 0.872 | 19157 |
| DYNMODPSO | 0.804 | 0.815 | 0.792 | 0.794 | 0.793 | 34038 |
| DYNMOGA | 0.747 | 0.761 | 0.766 | 0.763 | 0.769 | 15542 |
| Merging and Splitting | | | | | | |
| Algorithm | 1 | 2 | 3 | 4 | 5 | Time(s) |
| TMOGA | 0.902 | **0.918** | **0.917** | **0.945** | **0.916** | 6490(53.8) |
| DECS | 0.902 | 0.897 | 0.838 | 0.898 | 0.845 | 129074 |
| FacetNet | **0.952** | 0.914 | 0.915 | 0.913 | 0.902 | 10770 |
| DYNMODPSO | 0.832 | 0.823 | 0.784 | 0.749 | 0.748 | 37223 |
| DYNMOGA | 0.752 | 0.742 | 0.753 | 0.721 | 0.715 | 15565 |

Table 6. Clustering results and running time for large-scale data. Time in parentheses means the running time of feature transfer

## 7.5. Real-world Problem

In this part, we choose a real life problem to explore, the mobile phone communication network, and this practical problem does not give the real labels and *Modularity* is the indicator.

**Mobile Phone Communication Network**[2] The data set comes from cell phone calls located in the Isla Del Sueño area of Spain in June 2006, during which the selected 400 mobile phones and all the communications between them were recorded during 10 days. In the network, each node represents a mobile phone terminal, which is associated with an edge, indicating whether two mobile phone terminals were talking during this time.

**Result** In this real world problem, we can find TMOGA still achieves the best solutions compared with other 4 algorithms among all steps (see Fig. 8). We select $t = 6$ to observe the final communities structure, and the visualization is displayed in Fig. 9. Distinct from other datasets, the whole network is divided into several pieces and only remains local structures. As the edge relationship becomes more complex, it is tough to distinguish all edge relationship for DECS, so the result is not so desirable. FacetNet obtains ordinary result as we all expected. On the other hand, with the complex edge membership, random walk step is easier to migrate into other communities, thus the effect of random walk initialization from DYNMODPSO decreases with poor final results. DYNMOGA, is also stable enough on such complex networks and obtain the second greatest performance. For running time, TMOGA has the most excellent performance. On the other hand, since communities are more, the running time of FacetNet increases and is longer than TMOGA. DECS and DYNMODPSO, as expected, consume the longest computation time. DYNMOGA has similar time consumption as DYNMODPSO, due to the heavier computation of *Community Score* as the number of communities increases. More details about the running time of DYNMOGA will be discussed in 7.7. Therefore, TMOGA algorithm can obtains the best result in both solution quality and running time.

*7.6. Initialization Strategy*

As mentioned in [2], community detection problem is a N-P hard problem with rather complex solution space. So the quality of initial solutions in NSGA-II remarkably determines the final performance of algorithm. This part gives a comparable test for different initialization strategies and shows the state-of-art performance of feature transfer mechanism.

This test includes sequential initialization for dynamic networks. At each snapshot, we run each strategy to obtain 200 initial solutions by providing real label of last snapshot if needed, then calculate the average performance of the highest 20 solutions. Since different strategies can make a large difference in large-scale network, we choose the four dynamic networks in 7.4 as test data. *NMI* with real result is selected as criteria. The control strategies consist of 3 different strategies: random initialization from [3], label propagation with 5 iterations only, naïve feature transfer without label propagation.

---

[2] Data can be found at http://visualdata.wustl.edu/varepository/

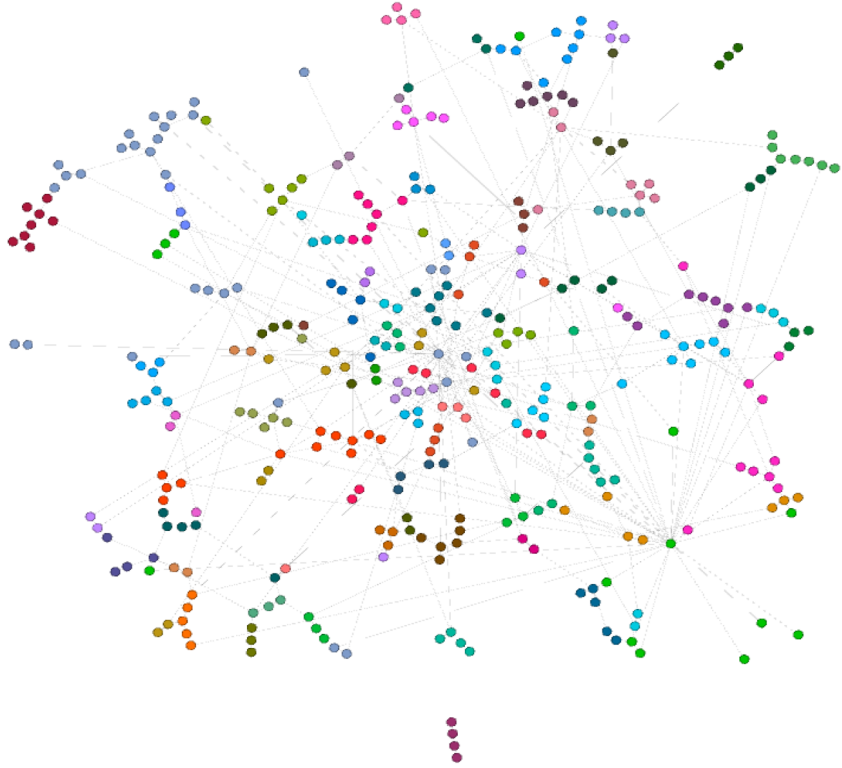

Fig. 9. Final cluster for TMOGA for Mobile Phone Communication at t = 6

**Result** As displayed in Fig.10, feature transfer mechanism can generate initial solutions with the highest quality among all the four networks. At $t = 1$, since there is no useful information, the performance is equal to label propagation strategy. When $t > 1$, feature transfer can improve about 5% performance to label propagation. On the other hand, since the member relationship of 20% nodes will be shuffled at each snapshot, the changes of network structures are significant. This phenomenon makes naïve feature transfer lose effectiveness. Thus label propagation that help adjust small cliques to be suitable for new network structure is necessary. Finally, random initialization has the lowest *NMI* as expected. It is worth mentioning that, even the initial solutions of feature transfer can have similar performance to the final solutions of DYNMODPSO, and better than the final solutions of DYNMOGA in 7.4. This experiment proves the state-of-art performance of feature transfer initialization.

## 7.7. Further Investigation on Some Modified Versions

To further explore the effect of TMOGA, we then propose two modified versions to make comparison. One is TMOGA2, where we use *Community Score* as snapshot cost and *Modularity* as selection criteria for Pareto solutions instead. Interestingly, this combination is what DYNMOGA chooses. The other version is TMOGA/SDE. It exploits Shift-based Density Estimation Method from [43] to recalculate crowding distances, which has better ability to maintain population diversity in high-dimensional multi-objective optimization. We test these algorithms on the four large-scale networks in 7.4 with *NMI* to ground truth as criteria.

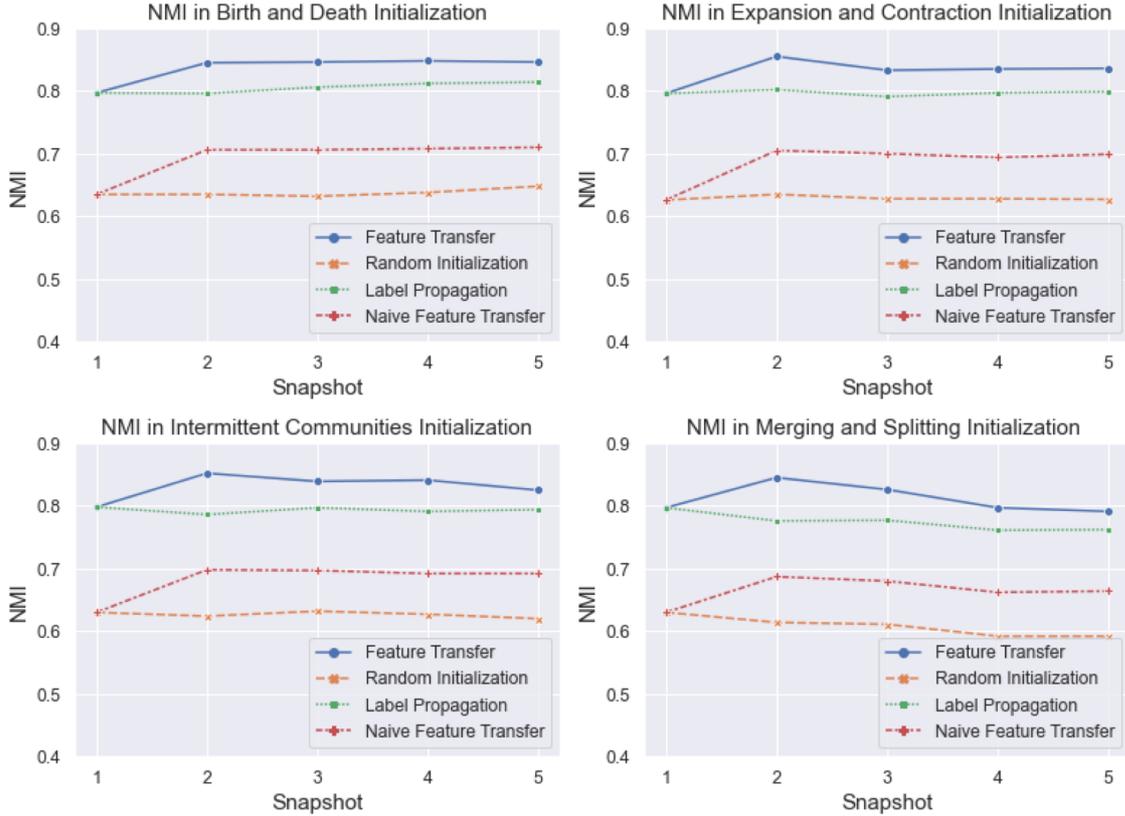

Fig. 10. Solution quality of different initialization strategies for large-scale data

**Result** From what has been shown in table 7, TMOGA has slightly better performance compared with TMOGA/SDE in these datasets. Although TMOGA/SDE is superior to TMOGA at every step in Intermittent Communities Model, but TMOGA has greater accuracy in the other datasets. It reveals the fact that Shift-based Density Estimation Method cannot significantly enhance the optimization process in low-dimensional multi-objective problem.

For TMOGA2, the results are notably worse than other 2 alternatives. To investigate the reason, we compare the final population of TMOGA and TMOGA2 at the 1st snapshot, to eliminate the influence of feature transfer and bias accumulation. As shown in figure 11, even if TMOGA2 obtains the solutions with higher *Community Score*, the *NMI* to ground truth is typically lower. However, by exploiting *Modularity* as snapshot cost, TMOGA gets better *NMI*, which indicates TMOGA has greater performance. To some extent, *Modularity* has positive correlation to *NMI* while *Community Score* has negative correlation on the contrary. Consequently, although some NSGA-II based algorithms (DYNMOGA, DYNMODPSO, etc.) use *Community Score* as snapshot cost objective, we strongly recommend to utilize *Modularity* instead.

| Birth and Death | | | | | | |
|---|---|---|---|---|---|---|
| Algorithm | 1 | 2 | 3 | 4 | 5 | Time(s) |
| TMOGA | **0.919** | **0.929** | **0.936** | 0.936 | 0.945 | 6103(52.3) |
| TMOGA2 | 0.796 | 0.806 | 0.827 | 0.829 | 0.824 | 23116(58.6) |
| TMOGA/SDE | 0.916 | 0.928 | 0.935 | **0.942** | **0.946** | 6274(59.1) |
| Expansion and Contraction | | | | | | |
| Algorithm | 1 | 2 | 3 | 4 | 5 | Time(s) |
| TMOGA | **0.927** | **0.933** | 0.934 | 0.944 | **0.971** | 6266(53.2) |
| TMOGA2 | 0.803 | 0.809 | 0.808 | 0.812 | 0.821 | 24164(61.4) |
| TMOGA/SDE | 0.906 | 0.928 | **0.937** | **0.946** | 0.948 | 6531(60.9) |
| Intermittent Communities | | | | | | |
| Algorithm | 1 | 2 | 3 | 4 | 5 | Time(s) |
| TMOGA | 0.914 | 0.925 | 0.932 | 0.948 | 0.946 | 5858(51.3) |
| TMOGA2 | 0.787 | 0.798 | 0.817 | 0.81 | 0.824 | 23599(58.6) |
| TMOGA/SDE | **0.916** | **0.937** | **0.946** | **0.959** | **0.952** | 5881(56.7) |
| Merging and Splitting | | | | | | |
| Algorithm | 1 | 2 | 3 | 4 | 5 | Time(s) |
| TMOGA | **0.902** | **0.918** | **0.917** | **0.945** | **0.916** | 6490(53.8) |
| TMOGA2 | 0.796 | 0.796 | 0.788 | 0.771 | 0.777 | 26300(62.9) |
| TMOGA/SDE | 0.899 | 0.908 | 0.912 | 0.902 | 0.905 | 7031(62.3) |

Table 7. Clustering results and running time for modified version of TMOGA. Time in parentheses means the running time of feature transfer.

For the running time, TMOGA is slightly shorter than TMOGA/SDE, while TMOGA2 is much longer. In NSGA-II algorithm, we need to calculate ($t*gen*3*pop$) times for snapshot cost, and only ($t*pop$) times for Pareto selection function. This huge difference can be applied to shorten the running time of algorithm if the calculation of snapshot cost is less complicated than the calculation of Pareto solutions selection function in large dataset. In our algorithm, the calculation of *Modularity* is much easier than *Community Score*, this is the reason why TMOGA has less running time compared with DYNMOGA, even extra time is needed for feature transfer.

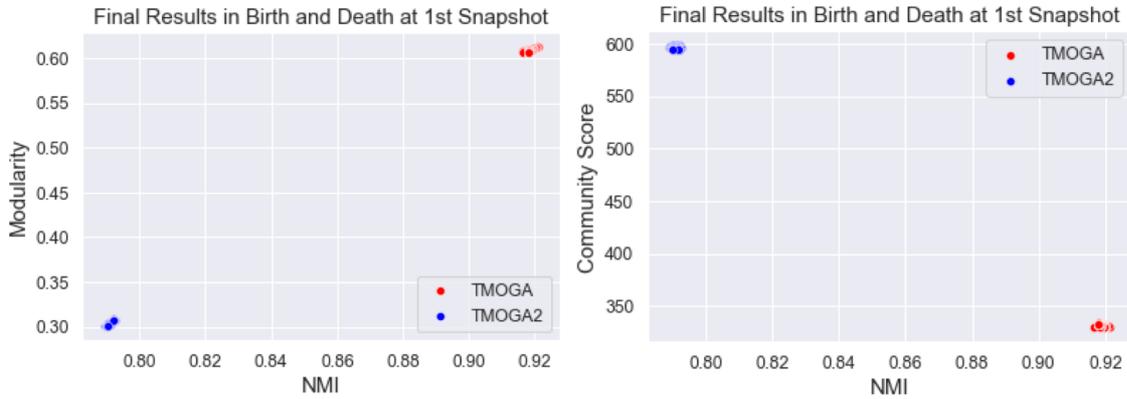

Fig. 11. Final population performance at 1$^{st}$ snapshot for large-scale data

## 8. Conclusion

Dynamic network community detection is an important problem of network science and artificial intelligence. At present, many scholars have proposed a series of algorithms to solve such problems, and many classic tests have been proposed to verify the effect of the algorithm. Few algorithms perform well in all of the tests. In view of the dynamic characteristics of dynamic network structure, this paper proposes a transfer learning based multi-objective genetic algorithm (TMOGA), which built on the traditional multi-objective algorithm, NSGA-II. For the first time, feature transfer in transfer learning is applied into this problem. We define cliques as features in communities, and extract the features and migrate them into initial population to enhance the genetic algorithm. Moreover, from an informational theoretical perspective, we build a theory to analyze dynamic network and apply the *Information Bottleneck* to prove the rationality of feature transfer which abstracts the community detection problems into the process of information compression. In addition, this paper designed a series of controlled trials. Small-scale benchmark data set proposed by Girvan and Newman, large-scale provided by Lin and Greene and mobile phone communication network, comparable test on different initialization strategies, additional experiment on two modified versions, are used to verify the operation effect of TMOGA algorithm. Experiments from these data sets have shown that TMOGA can always achieve the state-of-art effect on different dataset, by using feature transfer mechanism.

## Acknowledgements

## Appendix A. Parameters for generating large-scale dataset

In the large-scale problem, we test our algorithm on 4 types of datasets generated by the tools in [42]. To simulate different evolutionary events, there are some parameters to specify. The parameters for these 4 datasets are as follows:

| Model | Parameters | Value |
|---|---|---|
| Common Parameters | Number of nodes (-N) | 1000 |
| | Number of snapshots (-s) | 5 |
| | Average degree (-k) | 8 |
| | Maximum degree (-maxk) | 15 |
| | Mixing parameter (-muw) | 0.2 |
| | Minimum community size (-minc) | 24 |
| | Maximum community size (-maxc) | 35 |
| | Probability of node assignment (-p) | 0.2 |
| Birth and Death | Birth number (-birth) | 3 |
| | Death number (-death) | 3 |
| Expansion and Contraction | Expansion number (-expand) | 3 |
| | Contraction number (-contract) | 3 |
| | Expansion/contraction rate (-r) | 0.25 |
| Intermittent Communities | Hide rate (-hide) | 0.1 |
| Merging and Splitting | Merging number (-merge) | 3 |
| | Splitting number (-split) | 3 |

Table 8 Parameters for generating evolutionary event